# Evidence for a ferromagnetic quantum critical point in URhGe doped with Ru

N.T. Huy[a,*], D.E. de Nijs[a], A. Gasparini[a], J.C.P. Klaasse[a], A. de Visser[a], N.H. van Dijk[b]

[a]*Van der Waals-Zeeman Institute, University of Amsterdam, Valckenierstraat 65, 1018 XE Amsterdam, The Netherlands*
[b]*Department of Radiation, Radionuclides & Reactors, Delft University of Technology, Mekelweg 15, 2629 JB Delft, The Netherlands*

**Abstract**

We have investigated the evolution of ferromagnetic order in the correlated metal series $URh_{1-x}Ru_xGe$. Magnetization, transport and specific heat measurements provide convincing evidence for a ferromagnetic quantum critical point at the critical concentration $x_c = 0.38$. Here we report ac-susceptibility and magnetization measurements on selected samples with Ru doping concentrations near the critical point.

The intermetallic compound URhGe attracts much attention because itinerant ferromagnetism ($T_C = 9.5$ K) and unconventional superconductivity ($T_s = 0.25$ K) coexist [1]. Moreover, re-entrant superconductivity is observed for a magnetic field $B \approx 12$ T applied along the orthorhombic $b$-axis [1]. A natural explanation for the occurrence of these unusual superconducting states is that critical magnetic fluctuations stimulate the formation of spin-triplet Cooper pairs. Therefore it is of considerable interest to explore the magnetic properties of URhGe, especially as far as the ferromagnetic (FM) instability is concerned.

Recently, we have investigated the evolution of FM order in URhGe doped with Ru, Co and Si [2]. Magnetization and transport experiments on a series of polycrystalline $URh_{1-x}Ru_xGe$ samples down to $T = 2$ K showed that FM order, after an initial weak increase, is suppressed in a linear fashion and disappears at a critical concentration $x_c = 0.38$ (see Fig.1). In fact these experiments hinted at the existence of a FM quantum critical point (QCP) in the U(Rh,Ru)Ge system. This was corroborated by subsequent low temperature ($T \geq 0.3$ K) specific-heat and resistivity measurements [3]. At $x_c$ the phonon-corrected specific heat varies as $c \sim T\ln(T/T_0)$ over a wide $T$ range and $\chi(x)$ (*i.e.* $(c/T)|_{0.5K}$) goes through a pronounced maximum. The critical behaviour is also tracked by the exponent of the resistivity $\rho \sim T^n$, which attains a minimum value $n = 1.2$ at $x_c$. Taking into account that the magnetization data show a smooth suppression of the ordered moment $m_0(x)$, these results provide convincing evidence for a continuous FM quantum phase transition in the U(Rh,Ru)Ge system [3].

In this work we further investigate the phase diagram in the vicinity of the critical point. We report low temperature ($T \geq 0.23$ K) ac-susceptibility, $\chi_{ac}$, measurements in a field of $10^{-5}$ T. The data were taken on arc-melted polycrystalline samples annealed at 875 °C for 10 days. In Fig.2 we show the results for $x = 0.35$ and 0.375. The maxima in $\chi_{ac}$ signal FM order. For $x = 0.35$ we find $T_C = 1.0$ K. This value is somewhat smaller than the values deduced from the Arrott plot (1.3 K) [2] and from the specific heat (1.2 K) [3]. On the other hand $T_C = 0.4$ K for $x = 0.375$ nicely falls on the straight line with $dT_C/dx = -0.44$ K/at.%Ru, deduced from fitting the data for $x \geq 0.2$. The frequency ($f$) dependence of $\chi_{ac}$, investigated in the range 0.03-3.33 kHz, is weak as follows from the near overlap of the curves in Fig.2. $T_C(f)$ shows a small initial increase, but then saturates with $\Delta T_C/T_C \sim 5\%$ near the maximum frequency. The amplitude of $\chi_{ac}$ initially increases but is virtually constant for $f \geq 0.11$ kHz (see caption of Fig.2). These new $\chi_{ac}$ data are in line with the notion of a FM QCP at $x_c = 0.38$. Our $\chi_{ac}$ results manifestly differ from those recently reported for $x = 0.30$ [4] where a strong frequency dependence of $\chi_{ac}$ was taken as evidence for short-range rather than long-range magnetic order. Also the dc-magnetization results differ. Whereas in Ref.4 no spontaneous magnetization was detected for $x = 0.30$, we derive proper Arrott plots at least up to $x = 0.325$ (see Fig.1). We attribute these dissimilar results to a different sample quality, which may vary with the preparation method and annealing procedure.

In conclusion, we have investigated the evolution of magnetism in the correlated metal URhGe doped with Ru via magnetization, transport, specific heat, and ac-susceptibility measurements. Long-range FM order is suppressed and vanishes at a critical concentration $x_c = 0.38$. Magnetization and $\chi_{ac}$ data taken on samples in the vicinity of the critical point, provide additional evidence for long-range FM order for all $x < x_c$. The U(Rh,Ru)Ge alloys are unique in the sense that they present the first $f$-electron system with a FM QCP at ambient pressure (notice the vanishing of FM order in $CePd_{1-x}Rh_x$ is very much smeared [5]).

This work was supported by FOM (Dutch Foundation for Fundamental Research of Matter) and EU Cost P16 Action ECOM.

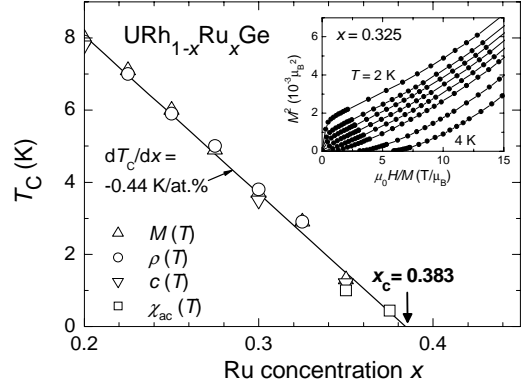

Fig.1 Curie temperature of $URh_{1-x}Ru_xGe$ alloys with $x \geq 0.2$ determined by different techniques as indicated. Open squares are from $\chi_{ac}$. The inset shows the Arrott plot for $x = 0.325$ with $T_C = 2.9$ K.

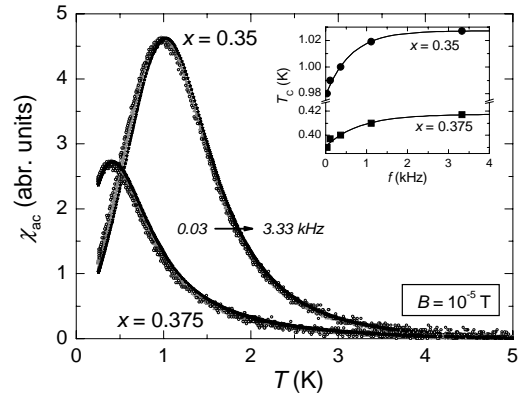

Fig. 2 Temperature dependence of the ac susceptibility in $URh_{1-x}Ru_xGe$ for $x = 0.35$ and 0.375 at frequencies $f = 0.03, 0.11, 0.35, 1.11$ and $3.33$ kHz. The data are scaled to the maximum amplitude of $\chi_{ac}$ at 3.33 kHz using multiplication factors of $1\pm0.03$ for $f = 0.11, 0.35$ and $1.11$ kHz and $\sim 1.2$ for $f = 0.03$ kHz. The inset shows $T_C(f)$.